\documentclass[conference]{IEEEtran}
\IEEEoverridecommandlockouts
\usepackage{cite}
\usepackage{amsmath,amssymb,amsthm,amsfonts,multicol, nccmath}
\usepackage{algorithm, multirow, ragged2e}
\usepackage[noend]{algpseudocode}
\usepackage{graphicx}
\usepackage{xcolor}
\definecolor{wV}{HTML}{08306b}
\definecolor{wV-DNN}{HTML}{323251}
\definecolor{woV}{HTML}{FFFFE0}

\usepackage{tikz}
\usepackage{pgfplots}
\pgfplotsset{compat=1.18}
\usetikzlibrary{patterns}
\usepackage[hidelinks]{hyperref}
\usepackage{subcaption}
\usepackage{float, array, tabularray, diagbox}
\UseTblrLibrary{diagbox}
\def\BibTeX{{\rm B\kern-.05em{\sc i\kern-.025em b}\kern-.08em
		T\kern-.1667em\lower.7ex\hbox{E}\kern-.125emX}}

\begin{document}
	
\title{Semi-Supervised Learning for Anomaly Detection in Blockchain-based Supply Chains}


\author{\IEEEauthorblockN{Do Hai Son\IEEEauthorrefmark{1}, Bui Duc Manh\IEEEauthorrefmark{2}, Tran Viet Khoa\IEEEauthorrefmark{3}, Nguyen Linh Trung\IEEEauthorrefmark{4},\\ Dinh Thai Hoang\IEEEauthorrefmark{2}, Hoang Trong Minh\IEEEauthorrefmark{5}, Yibeltal Alem\IEEEauthorrefmark{3}, and Le Quang Minh\IEEEauthorrefmark{1} \\}
    \IEEEauthorrefmark{1} VNU Information Technology Institute, Hanoi, Vietnam. \\
    \IEEEauthorrefmark{2} School of Electrical and Data Engineering, University of Technology Sydney, Australia. \\
    \IEEEauthorrefmark{3} University of Canberra, Australia. \\
    \IEEEauthorrefmark{4} AVITECH, VNU University of Engineering and Technology, Hanoi, Vietnam. \\
    \IEEEauthorrefmark{5} Posts and Telecommunications Institute of Technology, Vietnam.         
    \thanks{
    Corresponding author: Nguyen Linh Trung (linhtrung@vnu.edu.vn).}
}

\maketitle

\begin{abstract}
Blockchain-based supply chain (BSC) systems have tremendously been developed recently and can play an important role in our society in the future. In this study, we develop an anomaly detection model for BSC systems. Our proposed model can detect cyber-attacks at various levels, including the network layer, consensus layer, and beyond, by analyzing only the traffic data at the network layer. To do this, we first build a BSC system at our laboratory to perform experiments and collect datasets. We then propose a novel semi-supervised DAE-MLP (Deep AutoEncoder-Multilayer Perceptron) that combines the advantages of supervised and unsupervised learning to detect anomalies in BSC systems. The experimental results demonstrate the effectiveness of our model for anomaly detection within BSCs, achieving a detection accuracy of $96.5$\%. Moreover, DAE-MLP can effectively detect new attacks by improving the F1-score up to $33.1$\% after updating the MLP component.

\end{abstract}


\section{Introduction}\label{Intro}

Blockchain technology has experienced significant growth across various sectors, including business, finance, and beyond~\cite{ali2018applications}. Since the launch of the Ethereum network in 2016, blockchain's application has expanded well beyond cryptocurrency.
Nowadays, blockchain shows its impacts in many real-world applications, e.g., smart health, smart cities, and so on~\cite{ali2018applications}. This is largely due to its advanced properties, including immutability, transparency, and fault tolerance. However, despite these advantages, blockchain systems are not without vulnerabilities. These weaknesses can lead to severe consequences, such as financial theft and information breaches~\cite{networkattacks1}.

In this work, we consider a rising application of blockchain technology, i.e., supply chain management.
The blockchain-based supply chain, management, utilizes 
smart contracts (SMs) to record various processes including development, transportation, and consumption of commodities, such as agricultural products~\cite{Salah2019}.
Although BSCs offer numerous advantages, the inherent vulnerabilities of blockchain technology can lead to significant issues, such as counterfeit goods deceiving consumers~\cite{Vangala2021}. To deal with this problem, general intrusion detection system (IDS) techniques are applied to supply chain systems. In~\cite{Cheng2021}, the authors pointed out that IDS can be implemented at various levels within BSCs, including the network, consensus, and beyond layers.
It is also highlighted that deploying IDS at the network layer is crucial for the early detection of cyber-attacks on both traditional computer networks and blockchain networks. Therefore, in this paper, we focus on studying IDS methods to detect cyber-attacks at the network layer for BSCs.

Many studies have been implemented to develop IDS for supply chains on traditional computer networks~\cite{Asante2023}.
In these works, deep learning (DL) is a state-of-the-art technique in terms of cyber-attack detection. Because supervised learning needs to be trained by known cyber-attacks, DL-based anomaly detection can be classified into two methods, i.e., unsupervised learning and semi-supervised learning~\cite{dua2016data}. On the one hand, unsupervised learning uses unlabeled datasets to train the neural network (NN).
The advantage of unsupervised learning is that it can detect unknown cyber-attacks without knowing their characteristics~\cite{dua2016data}. In~\cite{Araujo2023}, the authors proposed a generative adversarial network (GAN)-based IDS to detect cyber-attacks on traditional computer networks. This approach can achieve accuracy up to 97\% for detecting the distributed denial-of-service (DDoS) in a well-known dataset. 
On the other hand, semi-supervised learning uses a part of labeled datasets to train an NN to enhance the performance for the anomaly detection of unsupervised learning. 
In~\cite{Dong2021}, the authors proposed a cascade structure of an auto-encoder (AE) and a deep neural network (DNN) to detect cyber-attacks in the NSL-KDD dataset. This network can classify known cyber-attacks in the dataset instead of just detecting abnormal network behaviors.
However, the above studies are geared toward cyber-attacks on traditional computer networks and are not yet related to blockchain-based systems. 

From the perspectives of IDS for blockchain-based systems, other works tried to use the above DL approaches for blockchain networks. In~\cite{Kim2022}, $11$ features of the Bitcoin network, e.g., \textit{version, average, max size}, and \textit{cost of messages}, are extracted to detect two types of blockchain cyber-attacks in the network layer, i.e., Eclipse and DoS. The proposed security mechanism can detect perfectly trained blockchain cyber-attacks using a typical AE network.
In~\cite{Sanjalawe2023}, the authors used a semi-supervised GAN approach to detect the anomaly in a public blockchain dataset for general purposes. The authors in~\cite{Sanjalawe2023} used $21$ transactions (txs)-based features of the dataset, e.g., \textit{gas price}, \textit{gas cost}, \textit{amount of Ethereum tokens}, and so on, to achieve accuracy up to 95\% for detecting anomalies in blockchain networks. In this work, we aim to detect cyber-attacks at the network layer in BSCs. 

From the above literature, anomaly detection at the network layer using DL in BSCs has to face various challenges. The first challenge is the lack of a cyber-attack dataset for BSCs. Recently, in~\cite{Khoa2022}, the authors introduced the first synthetic dataset for studying cyber-attacks in the Ethereum network. However, it is either not for BSC purposes or does not include cyber-attacks in the consensus and beyond layers~\cite{Cheng2021}.
Another challenge is the difficulty of detecting abnormal txs at the network layer without the labeled dataset.
For example, the overflow and underflow (OaU) vulnerability~\cite{networkattacks1}, in the consensus and beyond layers, leads to adversarial txs that are indistinguishable from honest txs to the same SM just by an unsupervised network.

To tackle the first challenge, we first built a BSC in the laboratory environment to capture the dataset. Our dataset consists of various types of cyber-attacks\footnote{Simulation codes and dataset are provided to reproduce the results in this paper: \nolinkurl{https://github.com/DoHaiSon/DAE-MLP}}: (i) at the network layer are brute-force password (BP) and denial of service (DoS), and (ii) at the consensus and beyond layers are DoS with block gas limit (DoS\_gas), OaU, and flooding of transactions (FoT). To collect traffic data on Ethereum nodes, we used the BC-ID tool in~\cite{Khoa2022} and the traceability protocol in~\cite{Salah2019}. 
To address the second challenge, we then propose a semi-supervised network named DAE-MLP to detect cyber-attacks in BSC systems. Different from~\cite{Dong2021} where the authors used a cascade structure in their neural network, we parallel the Deep AE (DAE) component (unsupervised) and multilayer perceptron~(MLP) component (supervised). The output of DAE-MLP is a combined anomaly score that is used to detect abnormal behaviors. 
There are two reasons for proposing this network structure. First, DAE-MLP can detect anomalies in the consensus and beyond layers with only the information from captured packets at the network layer, thanks to the labeled dataset, while retaining part of the advantage of an unsupervised learning-based detector.
Second, MLP's hyper-parameters can be individually updated for novel cyber-attacks without re-training the DAE component.
Our experiments show that our proposed DAE-MLP network can outperform other considered models (i.e., unsupervised learning networks and binary classifiers) with an accuracy of up to $96.5$\%. Additionally, DAE-MLP shows its ability to update novel cyber-attacks. F1-score of detecting new cyber-attacks before and after updating the MLP component is improved up to~$33.1$\%.


\section{BSC Systems and Our Proposed Anomaly Detection Model}\label{model}
\begin{figure}[t]
    \centering
    \includegraphics[width=\linewidth]{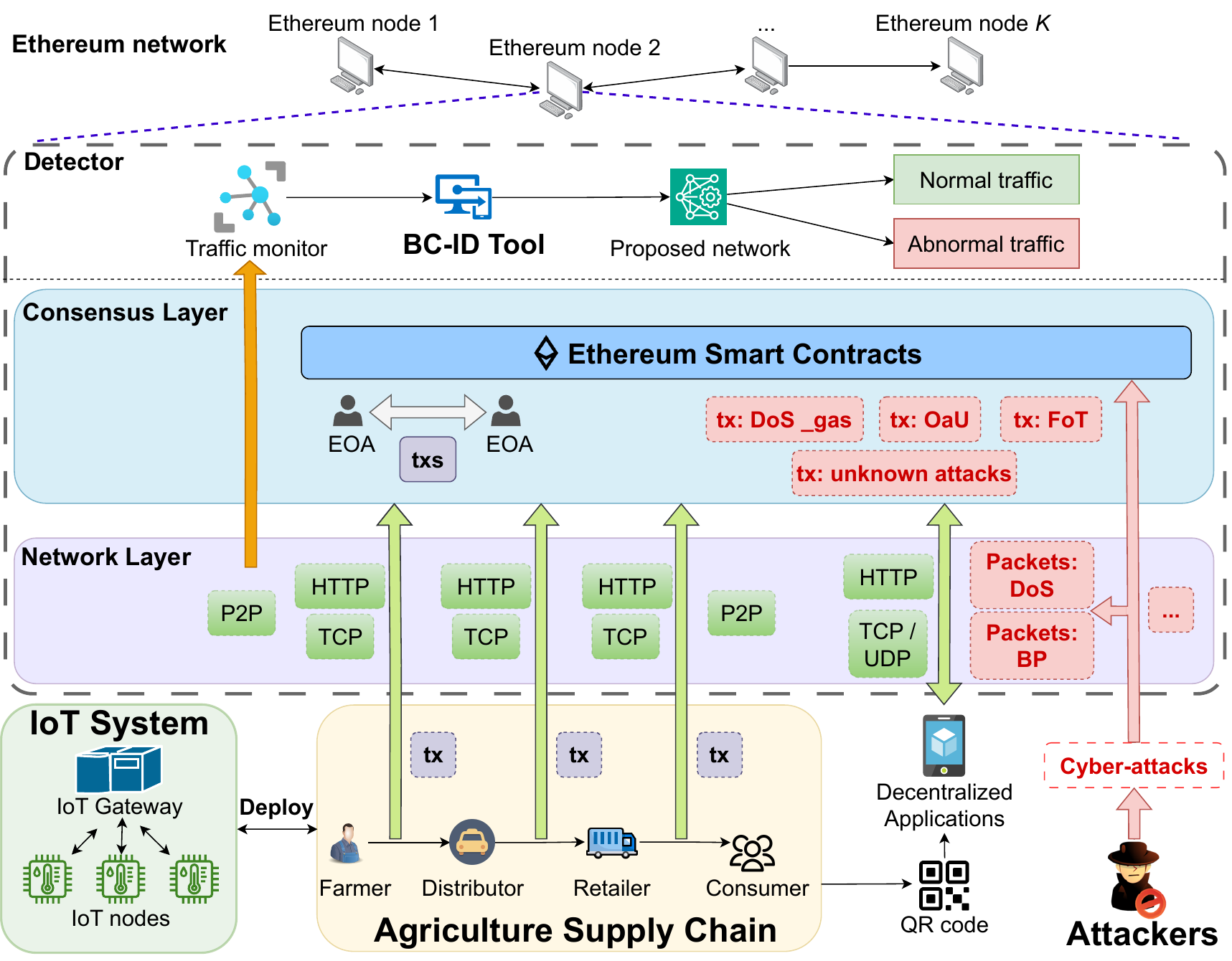}
    \caption{Proposed system model of anomaly detection on the Ethereum-based agriculture supply chain.}
    \label{fig:system}
    \vspace*{-0.5cm}
\end{figure}
In this paper, we study blockchain-based supply chains; an example of such applications is agriculture traceability. To do this, we first briefly present the protocol of a BSC system for agriculture traceability. We then demonstrate our system model to detect abnormal traffic data in the BSC at the network layer. 

\subsection{Blockchain-based Agriculture Supply Chains}
Since BSC is an emerging technology, it requires both modern hardware and communication protocols. Regarding hardware, at the farm field, BSC systems expect IoT devices, i.e., IoT sensors (e.g., humidity, temperature, motion, etc.), IoT gateways, and blockchain nodes (e.g., Ethereum). The IoT sensors read the environmental information and transmit it to IoT gateways. These values are packed into txs at the gateway and broadcast to the blockchain nodes based on BSC protocols. In this study, we use a BSC protocol that was proposed in~\cite{Salah2019}. In~\cite{Salah2019}, the authors introduced an SM and storage protocols for a soybean traceability process. In detail, there are seven participating entities in the traceability process, i.e., seed company, farmer, grain elevator, grain processor, distributor, retailer, and end customer. Each entity has its role, such as a seed company having to update the following information into the traceability SM, i.e., company name, lot coordinates, seed brand, certifying agency, and variety. Following that, farmers also have to push information, e.g., product ID, field ID, chemical application, harvest data, date sole, and so on, during cultivation. The rest of the entries add their information about the traceability process into the SM. Ultimately, the customers can use the product ID (e.g., encrypted in a QR code) to look up the BSC smart contract. The output encompasses the entire journey of the product, starting from its initial stage as a seed, through the growth and transportation processes, until it reaches the supermarket. This information is guaranteed to be reliable due to the immutability of blockchain technology. In this paper, we use four participating entities (i.e., farmer, distributor, retailer, and consumer) to collect data of BSC for analysis. The details are described in Fig.~\ref{fig:system}.



\subsection{Proposed Anomaly Detection Model}
Fig.~\ref{fig:system} shows our proposed system model for anomaly detection in BSCs. We assume that several participating entities and users interact with BSC through a blockchain end-point, i.e., Ethereum node. A detector is deployed at the top level of each Ethereum node. This detector comprises three components: a traffic monitor, a BC-ID tool, and our proposed DAE-MLP model.
First, the traffic monitor captures the network traffic data from network interfaces of the Ethereum node via `libpcap-dev' package. The captured traffic data consists of packets related to the network layer (e.g., ACK, SYN, etc.) and consensus and beyond layers (e.g., Ethereum tokens exchanges, BSCs' txs, etc.). Second, we use the BC-ID tool in~\cite{Khoa2022} to extract 21 important features from these captured packets.
At this stage, network packets are transformed into samples to be analyzed by DL networks. There are two phases in the cyber-attack detection of DL networks including offline training and online detection.
During the offline training phase, both unlabeled and labeled samples, including normal and anomaly states, are used to train our proposed DAE-MLP network. In the online detection phase, the trained model is deployed into the detector. Thereby, it predicts anomalies in the extracted samples from the BC-ID tool. A key feature of our proposed DAE-MLP network is the ability to update the MLP component's hyper-parameters with newly recognized and labeled cyber-attacks, thereby enhancing the detector's accuracy.

\section{Proposed DAE-MLP Network}\label{Method}
\begin{figure}
    \centering
    \includegraphics[width=\linewidth]{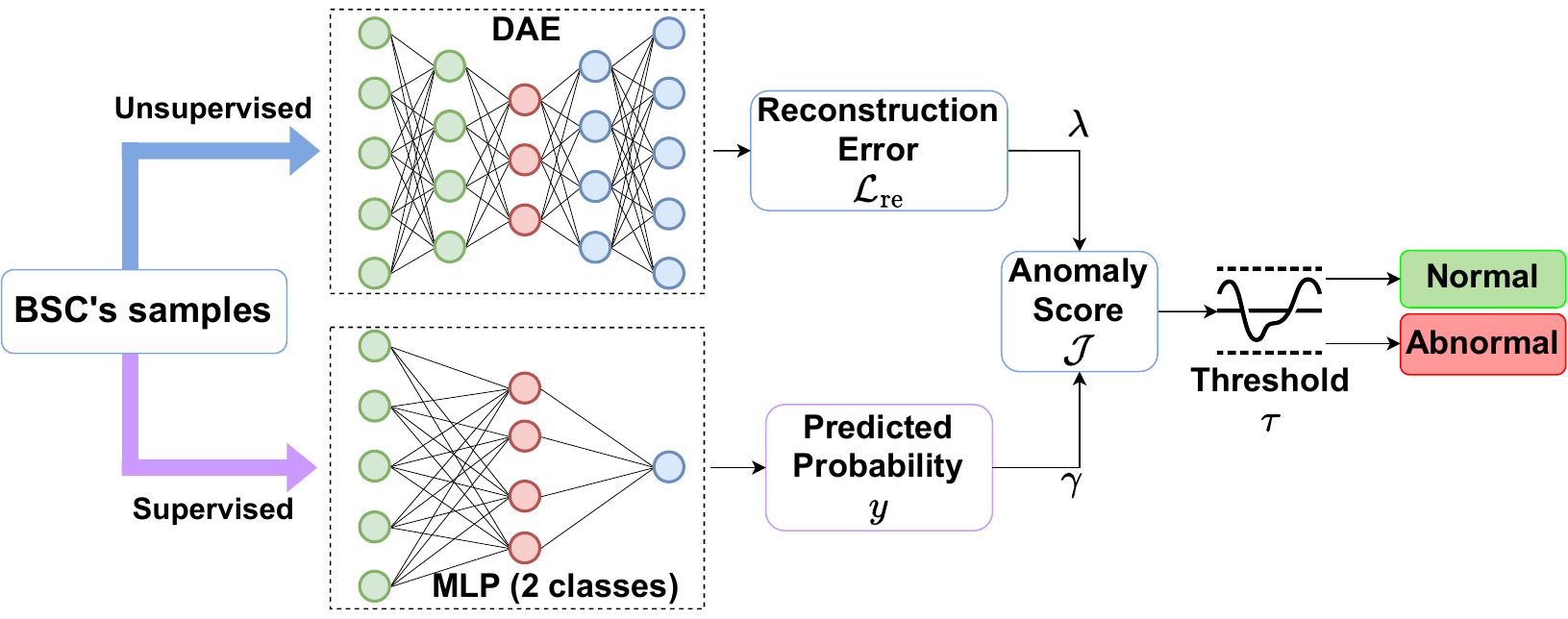}
    \caption{Proposed semi-supervised DAE-MLP network for anomaly detection in BSCs.}
    \label{fig:network}
    \vspace*{-0.5cm}
\end{figure}
Our proposed semi-supervised DAE-MLP network is illustrated in Fig.~\ref{fig:network}. The network consists of two main components, i.e., DAE and MLP networks~\cite{Schmidhuber2015}. The DAE nework is an unsupervised NN characterized by a typically narrow central layer, referred to as the ``bottleneck'', which is designed to reconstruct the input data at the output. Specifically, it consists of two main modules, i.e., the encoder~(E) and the decoder~(D). The encoder compresses the input data into a lower-dimensional representation, capturing essential features while reducing noise and redundancy. 
The $i$-th layer of this module is expressed as follows:
\begin{equation}
    \mathbf{z}(\mathbf{x})^{(i)} = f^{(i)}_{\text{E}} \left(\mathbf{W}^{(i)}_{\text{E}} \mathbf{x}^{(i-1)} + \mathbf{b}^{(i)}_{\text{E}} \right),
\end{equation}
for $i = 1,\ldots, I$. At the first layer, $\mathbf{x}^{(0)} \in \mathbb{R}^n$ is an input sample with $n$ features. $\mathbf{W}^{(1)}_{\text{E}} \in \mathbb{R}^{m \times n}$ and $\mathbf{b}^{(1)}_{\text{E}} \in \mathbb{R}^{m}$ are weight matrix and bias vector of encoder module, respectively. $f^{(i)}_{\text{E}}$ denotes the activation function, e.g., ReLU, Sigmoid, etc. The output $\mathbf{z}(\mathbf{x})^{(1)} \in \mathbb{R}^{m}$ is a latent representation which is utilized as the input of the next layer in this module. 
The decoder then reconstructs the latent space $\mathbf{z}(\mathbf{x})^{(I)}$ to form the output, which is expected the same as input.
Each layer of the decoder often increases the number of neurons. The $j$-th layer of this module is expressed as follows:

\begin{equation}
    \hat{\mathbf{x}}^{(j)} = f^{(j)}_{\text{D}} \left(\mathbf{W}^{(j)}_{\text{D}} \mathbf{z}(\mathbf{x})^{(j-1)} + \mathbf{b}^{(j)}_{\text{D}} \right),
\end{equation}
for $j=1,\ldots, J$. $\mathbf{W}_{\text{D}} \in \mathbb{R}^{n \times m}$ and $\mathbf{b}_{\text{D}} \in \mathbb{R}^{n}$ are weight matrix and bias vector of decoder module, respectively. The output $\hat{\mathbf{x}}^{(J)} \in \mathbb{R}^{n}$ is reconstructed version of $\mathbf{x}^{(0)}$.
The difference between the input and output, also referred to as reconstruction error (re), can be evaluated by the mean squared error (MSE), as follows:
\begin{equation}
    \mathcal{L}_{\text{re}} (\mathbf{x}^{(0)}, \hat{\mathbf{x}}^{(J)}) = \frac{1}{n}\sum_{a=1}^{n} (x_a^{(0)} - \hat{x}^{(J)}_a)^2.
\end{equation}

In our framework, the DAE is trained using normal traffic samples to learn the representation of the normal state. When network traffic containing cyber-attack samples is fed into the trained DAE model, the output significantly deviates from these inputs. This occurs because the distribution of abnormal samples differs from that of normal samples used during the DAE training process. Thus, $\mathcal{L}_{\text{re}} (\mathbf{x}^{(0)}, \hat{\mathbf{x}}^{(J)})$ of abnormal states will be higher than the loss observed in the normal state. By setting a threshold ($\tau$), we can effectively classify abnormal samples in network traffic of BSCs.

The latter component, a supervised MLP network (M), acts as a binary classifier (bi-classifier). The $k$-th layer, for $k=1,\ldots, K$, of bi-classifier is given as follows:
\begin{equation}
    \mathbf{y}(\mathbf{x})^{(k)} = f^{(k)}_{\text{M}} \left(\mathbf{W}^{(k)}_\text{M} \mathbf{x}^{(k-1)} + \mathbf{b}^{(k)}_{\text{M}} \right),
\end{equation}
where notations and their shape resemble those of a layer in the encoder of the DAE network, as mentioned above. However, at the last layer of MLP component, we transform the input $\mathbf{x}^{(K-1)}$ into a scalar value instead of a vector, i.e., $0 \le y^{(K)} \le 1$. This value represents the predicted probability. If $y^{(K)} \ge 0.5$, the label is assigned as $1$ (anomaly class); otherwise, the label is $0$ (normal class).

The DAE-MLP then utilizes both reconstruction loss from the DAE component and predicted probability from the MLP component to form a total anomaly score ($\mathcal{J}$), as follows:
\begin{equation}
    \mathcal{J} = \lambda \mathcal{L}_{\text{re}} + \gamma y,
\end{equation}
where $\lambda$ and $\gamma$ are weights of reconstruction loss and predicted probability, respectively. Nonetheless, DAE and MLP networks have two different loss functions (i.e., MSE and binary cross-entropy, respectively) and generate distinct loss values in each training epoch. Hence, we combine two loss values in an epoch into a final loss value ($\mathcal{L}_{\text{total}}$), to update hyper-parameters in both networks at once. This is achieved by weighting the losses with an $\alpha$ ratio.
\begin{equation}
    \begin{aligned}
        \mathcal{L}_{\text{total}} = \alpha \mathcal{L}_{\text{DAE}} + (1-\alpha) \mathcal{L}_{\text{MLP}}.
    \end{aligned}
\end{equation}

Based on the anomaly score, DAE-MLP decides whether the state of a sample, $A(\mathbf{x}, \lambda, \gamma, \tau, \alpha)$, is an anomaly or not by
\begin{equation}   
    \label{eq:threshold}
    A(\mathbf{x}, \lambda, \gamma, \tau, \alpha) = 
        \left\{\begin{matrix}
        &\text{Normal},  &\text{if} \quad \mathcal{J} \ge \tau, \\
        &\text{Anomaly}, &\text{if} \quad \mathcal{J} < \tau.
        \end{matrix}\right.
\end{equation}
\begin{algorithm}[!t]
\caption{An efficiency threshold for the DAE-MLP.
}
\label{algo:threshold}
\begin{algorithmic}[1]
    \Require{\par
    \hspace*{0.18cm}Initial threshold: $\tau \gets \tau_{\text{init}}$ \par
    Step size: $\zeta \gets 0.001$ \par
    Decay rate: $r \gets 0.5$ \par
    Decay times: $D \gets 15$ \par
    Number of worse accuracies in a row: $c \gets 0$ \par
    Maximum value of $c$: $c_{\text{max}} \gets 10$
    }
    \Ensure{\par
    \hspace*{-0.08cm}An efficiency threshold: $\tau$
    }
    \For{$d \gets 1$ \textbf{to} $D$}
        \While{True}
            \State $\text{Acc}_d \gets \texttt{evaluate\_model}(\lambda, \gamma, \tau)$
            \State $\tau \gets \tau + \zeta$

            \If{$\text{Acc}_d > \text{Acc}_{\text{best}}$}
                \State $\text{Acc}_{\text{best}} \gets \text{Acc}_d$
                \State $c \gets 0$
            \Else                
                \State $c \gets c + 1$
            \EndIf

            \If{$c = c_{\text{max}}$}
                \State $\tau = \tau - c_{\text{max}} \zeta$
                \State $\zeta \gets r \zeta$
                \State $c \gets 0$
                \State \textbf{break}
            \EndIf
        \EndWhile
    \EndFor
\end{algorithmic}
\end{algorithm}

As observed in Eq.~\eqref{eq:threshold}, threshold plays an essential role in anomaly detection accuracy using DAE-MLP~\cite{Kim2022}. Therefore, we design the algorithm~\ref{algo:threshold}, which is our small proposal to find an efficiency threshold for a trained DAE-MLP model to detect anomalies in BSCs. This algorithm is based on the grid search method with a decay rate. Initially, the threshold ($\tau_{\text{init}}$) value is a quantile function~\cite{Hyndman1996}, given by:
\begin{equation}
    \tau_{\text{init}} = Q_{\mathcal{L}_{\text{re}}\left(\mathbf{x}_{(\text{nor})}, \hat{\mathbf{x}}_{(\text{nor})}\right)}(\beta),
\end{equation}
where $\beta$ is the percentile of the quantile function. $\mathcal{L}_{\text{re}}\left(\mathbf{x}_{(\text{nor})}, \hat{\mathbf{x}}_{(\text{nor})}\right)$ is the reconstruction error of only normal state samples. The \texttt{evaluate\_model} function returns the accuracy value of trained DAE-MLP with input arguments, i.e., $\lambda, \gamma,$ and $\tau$. Note that, after updating the MLP component, algorithm~\ref{algo:threshold} must be executed again to adjust for the new hyper-parameters.
\section{Experimental Results}\label{SR}
\subsection{Experiment Setup and Evaluation Methods}
\begin{table}
\centering
\caption{Architecture of the proposed DAE-MLP.}
\label{tab:arc}
\resizebox{\linewidth}{!}{%
\begin{tabular}{|c|c|c|c|c|c|} 
\hline
\textbf{Component} & \multicolumn{1}{>{\centering}m{0.12\linewidth}|}{\textbf{Module}} & \multicolumn{1}{>{\centering}m{0.1\linewidth}|}{\textbf{Layer}} & \multicolumn{1}{>{\centering}m{0.15\linewidth}|}{\textbf{Input features}} & \multicolumn{1}{>{\centering}m{0.15\linewidth}|}{\textbf{Output features}} & \multicolumn{1}{>{\centering\arraybackslash}m{0.2\linewidth}|}{\textbf{Activation function}} \\ 
\hline
\multirow{7}{0.146\linewidth}{\Centering{}\textbf{DAE}} & \multirow{4}{0.112\linewidth}{Encoder} & Linear & 21 & 21 & ReLU \\ 
\cline{3-6}
 &  & Linear & 64 & 32 & ReLU \\ 
\cline{3-6}
 &  & Linear & 32 & 16 & ReLU \\ 
\cline{2-6}
 & \multirow{3}{0.112\linewidth}{Decoder} & Linear & 16 & 32 & ReLU \\ 
\cline{3-6}
 &  & Linear & 32 & 64 & ReLU \\ 
\cline{3-6}
 &  & Linear & 64 & 21 & Sigmoid \\ 
\hline
\multirow{3}{0.146\linewidth}{\Centering{}\textbf{MLP}} & \multirow{3}{0.112\linewidth}{Classifier} & Linear & 21 & 32 & ReLU \\ 
\cline{3-6}
 &  & Linear & 32 & 16 & ReLU \\ 
\cline{3-6}
 &  & Linear & 16 & 1 & Sigmoid \\
\hline
\end{tabular}
}
\end{table}
\begin{table}
\centering
\caption{Descriptions of dataset and training parameters.}
\label{tab:param}
\begin{tblr}{
  width = \linewidth,
  colspec = {Q[400]>{\centering\arraybackslash}Q[138]>{\centering\arraybackslash}Q[250]>{\centering\arraybackslash}Q[129]},
  row{1} = {c},
  hlines,
  vlines,
}
\textbf{Parameter} & \textbf{Value} & \textbf{Parameter} & \textbf{Value}\\
No. \textit{Normal} samples & $600,\!000$ & $\lambda$ & $0.5$\\
No. \textit{BP} samples & $25,\!293$ & $\gamma$ & $0.5$\\
No. \textit{DoS} samples & $100,\!000$ & $\alpha$ & $0.5$\\
No. \textit{DoS\_gas} samples & $91,\!128$ & Train / test ratio & $0.8/0.2$\\
No. \textit{OaU} samples & $50,\!998$ & Learning rate & $0.01$\\
No. \textit{FoT} samples & $100,\!000$ & $\beta$ & $0.9$
\end{tblr}
\vspace*{-0.5cm}
\end{table}
\begin{figure*}[!t]
    \centering
    \resizebox{\textwidth}{!}{%
    \begin{tikzpicture}
      \begin{axis}[
        xlabel={},
        ylabel={Percentage},
        set layers,
        ybar=5pt,
        x=6cm,
        bar width=10pt,
        symbolic x coords={Accuracy, Precision, Recall},
        grid,
        yminorgrids=true,
        ymin=0, ymax=105,
        enlarge x limits={0.3},
        xtick = {Accuracy, Precision, Recall},
        legend cell align=left,
        nodes near coords,
        nodes near coords style = {font=\fontsize{6}{6}\selectfont},
        ylabel near ticks, ylabel shift={-5pt},
        legend style={at={(160pt, -60pt)},anchor=south,draw=none, style={column sep=0.2cm}},
        minor tick num=3,
        legend columns=4,
        xlabel style={font=\bfseries},
        ylabel style={font=\bfseries},
        xtick pos=left, xtick style={draw=none},
        tick label style={font=\bfseries}
        ]
        \addplot[black,fill=woV,postaction={pattern=north east lines}] coordinates {
          (Accuracy, 51.40) (Precision, 51.40) (Recall, 56.35)
        };
        \addlegendentry{Sparse AE}
        \addplot[black,fill=red,postaction={pattern=grid}] coordinates {
          (Accuracy, 51.4) (Precision, 51.4) (Recall, 55.3)
        };
        \addlegendentry{DAE}
        \addplot[black,fill=wV-DNN,postaction={pattern=north west lines}, pattern color=white] coordinates {
          (Accuracy, 54.5) (Precision, 30.5) (Recall, 58.6)
        };
        \addlegendentry{GAN}
        \addplot[black,fill=blue!80!black,postaction={pattern=horizontal lines}, pattern color=white] coordinates {
          (Accuracy, 54.7) (Precision, 49.3) (Recall, 55.3)
        };
        \addlegendentry{One-class SVM}
        \addplot[black,fill=pink,postaction={pattern=crosshatch}, pattern color=black] coordinates {
          (Accuracy, 90.3) (Precision, 82.8) (Recall, 97.4)
        };
        \addlegendentry{Logistic Regression (bi-classifier)}
        \addplot[black,fill=orange,postaction={pattern=bricks}, pattern color=white] coordinates {
          (Accuracy, 90.3) (Precision, 80.7) (Recall, 99)
        };
        \addlegendentry{Multi-class SVM (bi-classifier)}
        \addplot[black,fill=brown,postaction={pattern=vertical lines}, pattern color=white] coordinates {
          (Accuracy, 91.2) (Precision, 90.3) (Recall, 91.9)
        };
        \addlegendentry{DBN (bi-classifier)}
        \addplot[black,fill=cyan,postaction={pattern=sixpointed stars}, pattern color=white] coordinates {
          (Accuracy, 92.1) (Precision, 87.6) (Recall, 96.2)
        };
        \addlegendentry{KNN (bi-classifier)}
        \addplot[black,fill=wV,postaction={pattern=dots}, pattern color=white] coordinates {
          (Accuracy, 96.5) (Precision, 96.5) (Recall, 96.3)
        };
        \addlegendentry{Proposed DAE-MLP}
      \end{axis}
    \end{tikzpicture}
    }
    \caption{The performance of proposed DAE-MLP vs. other anomaly detectors and bi-classifiers.}
    \label{fig:perf}
\end{figure*}
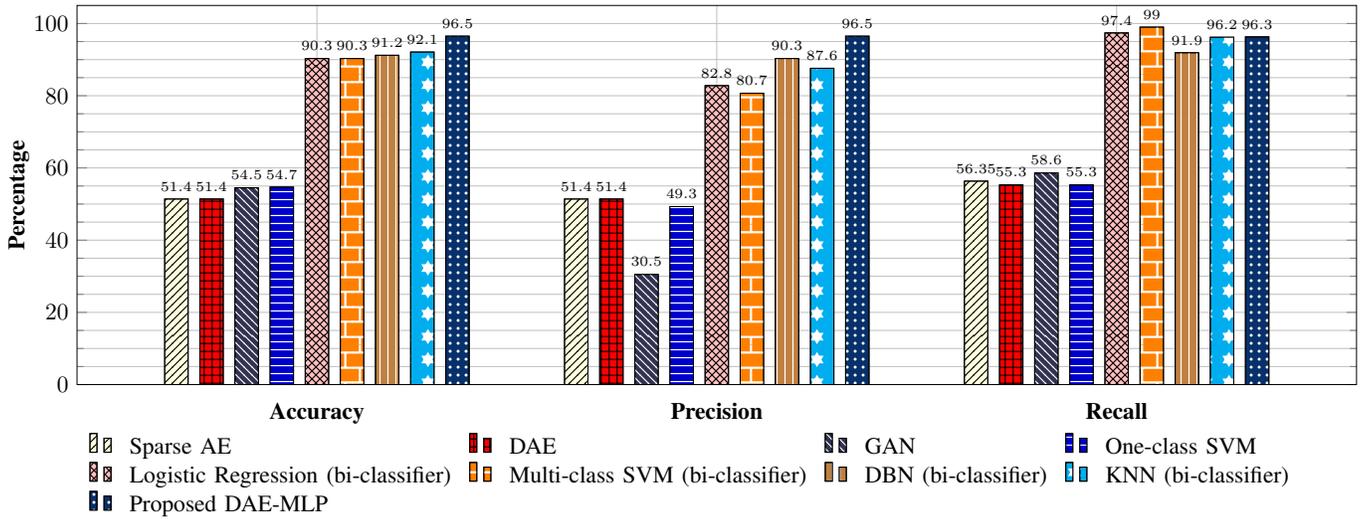
Real-world blockchain cyber-attacks aim at both peer-to-peer systems (i.e., network layer) and blockchain applications (i.e., consensus and beyond layers)~\cite{networkattacks1}. These attacks caused various consequences such as delays in block confirmation time, loss of digital assets, and fake information from the past. In this paper, we perform experiments to replicate five types of cyber-attacks that can directly impact BSCs as follows:
\begin{itemize}
    \item Cyber-attacks in the network layer~\cite{Khoa2022}:
    \begin{itemize}
        \item \textit{Brute-force Password (BP)}: Hackers target stealing EOA private keys of participating entities and users in BSCs using the BP attack.
        \item \textit{Denial of Service (DoS)}: Attackers send millions of network packets to blockchain nodes to increase the response time of BSC. 
    \end{itemize}

    \item Cyber-attacks in the consensus and beyond layers~\cite{networkattacks1, Cheng2021}:
    \begin{itemize}
        \item \textit{DoS with Block Gas limit (DoS\_gas)}: Functions inside the SM of BSCs could be temporarily unavailable if the required gas to execute these functions exceeds the gas limit. Hackers exploit this weakness to disrupt BSC's services.
        \item \textit{Overflow and Underflow (OaU)}: OaU vulnerabilities may be present in condition-checking statements in programming, potentially allowing hackers to bypass these checks and insert incorrect information into BSCs.
        \item \textit{Flooding of Transactions (FoT)}: Attackers send massive meaningless txs to blockchain nodes to increase the response time of BSC, similar to DoS in the network layer.  
    \end{itemize}
\end{itemize}

In our laboratory, we launch an Ethereum network and a few servers as IoT gateways for generating datasets. 
The servers perform as participating entities in the BSC. They also act as users to update and track the information in this system. Additionally, we use another server to perform five types of cyber-attacks on the Ethereum network and the SM of the BSC. We then use the BC-ID tool in~\cite{Khoa2022} to collect and extract features from network traffic of the blockchain nodes to create a dataset. Table~\ref{tab:arc} shows the architecture of our proposed DAE-MLP that will be used in the experiments in this paper. Table~\ref{tab:param} provides descriptions of our collected dataset within the BSC system. This table also provides information about weights and other parameters during the DAE-MLP training process. 
In this paper, we use several metrics to evaluate models, i.e., accuracy, F1-score, precision, and recall. These metrics are estimated using built-in functions from a widely-used machine learning library, i.e., scikit-learn~\cite{scikit}. Additionally, precision and recall metrics are calculated using the `macro' averaging method to ensure fairness among classes with different sample sizes.
\subsection{Performance Evaluation}
\subsubsection{Performance Comparison}
Fig.~\ref{fig:perf} provides the performance results in accuracy, precision, and recall of our proposed DAE-MLP network and other models, such as unsupervised anomaly detectors, and several binary supervised models (bi-classifiers).
In Fig.~\ref{fig:perf}, we can see that the accuracies in cyber-attack detection of the unsupervised networks, i.e., Sparse DAE, DAE~\cite{IERACITANO202051}, GAN, and one-class SVM have low accuracy values, i.e., $51.4, 51.4, 54.5$, and $54.7$, respectively. These results demonstrate that unsupervised networks are not effective in detecting cyber-attacks in the consensus and beyond layers by analyzing the network layer's packets in our dataset. Besides, supervised bi-classifiers, i.e., Logistic Regression, Multi-class SVM, DBN~\cite{Khoa2022}, and KNN, can achieve outstanding accuracy values, i.e., $90.3, 90.3, 91.2$, and $92.1$, respectively. Based on these results, we can see considerable improvements in the accuracies from unsupervised anomaly detectors to bi-classifiers. However, our proposed (semi-) supervised DAE-MLP network can achieve performance in accuracy, precision, and recall of about $96.5, 96.5$, and $96.3$, respectively. By employing the supervised component (i.e., MLP), the DAE-MLP can obtain higher performance results compared to other considered models
Overall, the accuracy, precision, and recall of our proposed DAE-MLP network outperform other considered models for anomaly detection in our dataset.

\begin{table}
\centering
\caption{Performance of proposed DAE-MLP with different threshold values.}
\label{tab:threshold}
\begin{tblr}{
  width = \linewidth,
  colspec = {Q[100]Q[71]Q[71]Q[71]Q[71]Q[71]Q[71]},
  column{2-7} = {c},
  hlines,
  vlines,
}
\diagbox[width=\dimexpr\linewidth+2\tabcolsep, height=1.1cm]{}{${\tau}$} & {${\mu}$~\cite{Najari2021}} & ${0.8\mu}$ & ${0.5\mu}$ & ${0.2\mu}$ & $\tau$~\cite{Homayouni2020} & ${\tau}_{\text{ours}}$\\
\textbf{Accuracy} & 43.67 & 43.62 & 45.40 &  94.74 & 95.45 & \textbf{96.52}\\
\textbf{Precision} & 52.94 & 52.87 & 54.09 & 93.59 & 96.10 & \textbf{96.47}\\
\textbf{Recall} & 67.22 & 66.63 & 64.19 & 95.66 & 94.53 & \textbf{96.29}
\end{tblr}
\vspace*{-0.5cm}
\end{table}

\subsubsection{Threshold Values}
Table~\ref{tab:threshold} provides the performance of DAE-MLP with different threshold values. We compare our proposed threshold value with those in~\cite{Najari2021} and~\cite{Homayouni2020}. In~\cite{Najari2021}, the threshold is a value within the range from $\mathcal{L}_{\text{re}}\left(\mathbf{x}_{(\text{nor})}, \hat{\mathbf{x}}_{(\text{nor})}\right)$ to $\mathcal{L}_{\text{re}}\left(\mathbf{x}_{(\text{ano})}, \hat{\mathbf{x}}_{(\text{ano})}\right)$, where $\mathcal{L}_{\text{re}}\left(\mathbf{x}_{(\text{ano})}, \hat{\mathbf{x}}_{(\text{ano})}\right)$ is the reconstruction error of only anomaly state samples. Table~\ref{tab:threshold} considers several values in this range, i.e., mean ($\mu$), $0.8\mu$, $0.5\mu$, and $0.2\mu$. On the other hand,~\cite{Homayouni2020} proposed to used $\tau = Q_{\mathcal{L}_{\text{re}}\left(\mathbf{x}_{(\text{nor})}, \hat{\mathbf{x}}_{(\text{nor})}\right)}(0.9)$ as the threshold value. In this work, we use this value $\tau$ as the initial value $\tau_{\text{init}}$ for algorithm~\ref{algo:threshold}. In general, we can observe that our threshold value, $\tau_{\text{ours}}$, gives more precision than other values with an accuracy of up to $96.5$\%. In contrast, the threshold values from~\cite{Homayouni2020} vary greatly, depending on the corresponding threshold values. Among these values, $0.8\mu$ and $0.2\mu$ are the worst and best threshold values with their accuracy, i.e., $43.62$\% and $94.74$\%, respectively. The reason can come from the significant overlap in the distribution of reconstruction errors between normal and anomaly state samples. 
Although the approach from~\cite{Homayouni2020} gives the acceptable accuracy of up to $95.45$\%, our proposed algorithm can enhance the performance of cyber-attack detection of the model. Specifically, with the initial threshold $\tau_{\text{init}}$, 
our proposed algorithm can provide better performance in accuracy and recall than that of~\cite{Homayouni2020} by approximately $1$\% and $2$\%, respectively.

\begin{table*}
\centering
\caption{Accuracy detection of novel cyber-attacks before and after updating the MLP component.}
\label{tab:update}
\begin{tabular}{|>{\centering}m{0.15\linewidth}|>{\centering}m{0.05\linewidth}|>{\centering}m{0.05\linewidth}|>{\centering}m{0.05\linewidth}|>{\centering}m{0.05\linewidth}|>{\centering}m{0.05\linewidth}|>{\centering}m{0.05\linewidth}|>{\centering}m{0.05\linewidth}|>{\centering}m{0.05\linewidth}|>{\centering}m{0.05\linewidth}|>{\centering\arraybackslash}m{0.05\linewidth}|} 
\hline
\diagbox[width=\dimexpr\linewidth+2\tabcolsep]{}{\vspace{.2cm}\hspace*{\linewidth}\textbf{Missing class}} & \multicolumn{2}{>{\centering}m{0.104\linewidth}|}{\textbf{BP}} & \multicolumn{2}{>{\centering}m{0.104\linewidth}|}{\textbf{DoS}} & \multicolumn{2}{>{\centering}m{0.104\linewidth}|}{\textbf{DoS\_gas}} & \multicolumn{2}{>{\centering}m{0.104\linewidth}|}{\textbf{OaU}} & \multicolumn{2}{>{\centering\arraybackslash}m{0.104\linewidth}|}{\textbf{FoT}} \\ 
\hline
\diagbox[width=\dimexpr\linewidth+2\tabcolsep]{}{\textbf{Case}} & \textbf{w/o} & \textbf{w} & \textbf{w/o} & \textbf{w} & \textbf{w/o} & \textbf{w} & \textbf{w/o} & \textbf{w} & \textbf{w/o} & \textbf{w} \\ 
\hline
\textbf{F1-score} & 74.03 & 79.39 & 60.98 & 94.08 & 86.08 & 91.80 & 81.68 & 86.79 & 88.70 & 93.18 \\ 
\hline
\textbf{Precision} & 64.18 & 66.74 & 79.69 & 88.82 & 86.56 & 87.85 & 78.18 & 79.92 & 88.10 & 89.02 \\ 
\hline
\textbf{Recall} & 87.46 & 97.95 & 49.38 & 100.0 & 85.60 & 96.11 & 85.51 & 94.96 & 89.30 & 97.76 \\
\hline
\end{tabular}
\end{table*}
\subsubsection{Adaptation}
In this session, we simulate a practical scenario where attacks are not previously trained by the model. We focus on the capacity to adapt to novel cyber-attacks of DAE-MLP network within BSCs. In particular, we assume that the training dataset lacks a type of attack, e.g., BP attack. However, this attack still appears while the model performs cyber-attack detection in BSCs. In this case, the model has to detect anomalies without being trained by them. After that, we record the anomaly and update the training dataset to improve the accuracy of cyber-attack detection. Table~\ref{tab:update} shows performance in F1-score, precision, and recall of the model when being trained without a type of attack (w/o) and being updated with the attack (w). In general, as in Table~\ref{tab:update}, without updating new attacks, the performance of the model reduces little with all types of attacks. In detail, our proposed model can detect FoT in the case of w/o with $88.7$\% and $88.1$\% in F1-score and precision, respectively. The results are slightly lower than the case of w with $93.18$\% and $89.02$\% in F1-score and precision, respectively. Besides, the proposed model can detect DoS attacks in the case of w/o with an F1-score of $60.98$\% and precision of $79.69$\%. In the case of w, the F1-score is $33.1$\% higher and the precision is $9.13$\% higher than the case of w/o, resulting in an F1-score of $94.08$\% and precision of $88.82$\%.
As a result, our proposed model can not only detect anomalies without previously being trained by the types of attacks but also improve the performance of detecting unknown attacks by allowing the model to update the detected cyber-attacks.  

\section{Conclusion}

In this paper, we proposed a novel semi-supervised learning framework for anomaly detection in BSCs. Specifically, we first implemented an Ethereum network in our laboratory to collect a dataset for blockchain-based supply chains, which includes normal and attack traffic data. 
After that, we designed DAE-MLP which is a semi-supervised model that can leverage the detection performance of the unsupervised DAE model by combining it with the supervised MLP model. 
Additionally, a grid search algorithm was employed to find the efficient threshold, thereby improving anomaly detection accuracy in the considered model.
The simulation results showed that our proposed model could have better performance in detecting anomalies than other supervised and unsupervised approaches. Moreover, the proposed model can not only detect anomalies without learning the type of attack but also improve the cyber-attack detection performance after updating new attacks to the model. 
Regarding future work, we aim to study more types of anomalies and design more intelligent models to enhance the accuracy of anomaly detection in BSCs.

	
\bibliographystyle{IEEEtran}
\bibliography{library.bib}	

\begin{thebibliography}{10}
\providecommand{\url}[1]{#1}
\csname url@samestyle\endcsname
\providecommand{\newblock}{\relax}
\providecommand{\bibinfo}[2]{#2}
\providecommand{\BIBentrySTDinterwordspacing}{\spaceskip=0pt\relax}
\providecommand{\BIBentryALTinterwordstretchfactor}{4}
\providecommand{\BIBentryALTinterwordspacing}{\spaceskip=\fontdimen2\font plus
\BIBentryALTinterwordstretchfactor\fontdimen3\font minus \fontdimen4\font\relax}
\providecommand{\BIBforeignlanguage}[2]{{%
\expandafter\ifx\csname l@#1\endcsname\relax
\typeout{** WARNING: IEEEtran.bst: No hyphenation pattern has been}%
\typeout{** loaded for the language `#1'. Using the pattern for}%
\typeout{** the default language instead.}%
\else
\language=\csname l@#1\endcsname
\fi
#2}}
\providecommand{\BIBdecl}{\relax}
\BIBdecl

\bibitem{ali2018applications}
M.~S. Ali, M.~Vecchio, M.~Pincheira, K.~Dolui, F.~Antonelli, and M.~H. Rehmani, ``{Applications of blockchains in the Internet of Things: A comprehensive survey},'' \emph{IEEE Communications Surveys \& Tutorials}, vol.~21, no.~2, pp. 1676--1717, Dec. 2018.

\bibitem{networkattacks1}
M.~Saad, J.~Spaulding, L.~Njilla, C.~Kamhoua, S.~Shetty, D.~Nyang, and D.~Mohaisen, ``{Exploring the attack surface of blockchain: A comprehensive survey},'' \emph{IEEE Communications Surveys \& Tutorials}, vol.~22, no.~3, pp. 1977--2008, Mar. 2020.

\bibitem{Salah2019}
K.~Salah, N.~Nizamuddin, R.~Jayaraman, and M.~Omar, ``Blockchain-based soybean traceability in agricultural supply chain,'' \emph{IEEE Access}, vol.~7, pp. 73\,295--73\,305, May 2019.

\bibitem{Vangala2021}
A.~Vangala, A.~K. Das, N.~Kumar, and M.~Alazab, ``Smart secure sensing for iot-based agriculture: Blockchain perspective,'' \emph{IEEE Sensors Journal}, vol.~21, no.~16, pp. 17\,591--17\,607, Aug. 2021.

\bibitem{Cheng2021}
J.~Cheng, L.~Xie, X.~Tang, N.~Xiong, and B.~Liu, ``A survey of security threats and defense on blockchain,'' \emph{Multimedia Tools and Applications}, vol.~80, pp. 30\,623--30\,652, Aug. 2021.

\bibitem{Asante2023}
M.~Asante, G.~Epiphaniou, C.~Maple, H.~Al-Khateeb, M.~Bottarelli, and K.~Z. Ghafoor, ``Distributed ledger technologies in supply chain security management: A comprehensive survey,'' \emph{IEEE Transactions on Engineering Management}, vol.~70, no.~2, pp. 713--739, Feb. 2023.

\bibitem{dua2016data}
S.~Dua and X.~Du, \emph{{Data Mining and Machine Learning in Cybersecurity}}.\hskip 1em plus 0.5em minus 0.4em\relax CRC Press, Apr. 2016.

\bibitem{Araujo2023}
P.~F. de~Araujo-Filho, M.~Naili, G.~Kaddoum, E.~T. Fapi, and Z.~Zhu, ``Unsupervised gan-based intrusion detection system using temporal convolutional networks and self-attention,'' \emph{IEEE Transactions on Network and Service Management}, vol.~20, no.~4, pp. 4951--4963, Dec. 2023.

\bibitem{Dong2021}
S.~Dong, Y.~Xia, and T.~Peng, ``Network abnormal traffic detection model based on semi-supervised deep reinforcement learning,'' \emph{IEEE Transactions on Network and Service Management}, vol.~18, no.~4, pp. 4197--4212, Dec. 2021.

\bibitem{Kim2022}
J.~Kim, M.~Nakashima, W.~Fan, S.~Wuthier, X.~Zhou, I.~Kim, and S.-Y. Chang, ``A machine learning approach to anomaly detection based on traffic monitoring for secure blockchain networking,'' \emph{IEEE Transactions on Network and Service Management}, vol.~19, no.~3, pp. 3619--3632, Sept. 2022.

\bibitem{Sanjalawe2023}
Y.~K. Sanjalawe and S.~R. Al-E’mari, ``Abnormal transactions detection in the ethereum network using semi-supervised generative adversarial networks,'' \emph{IEEE Access}, vol.~11, pp. 98\,516--98\,531, Sept. 2023.

\bibitem{Khoa2022}
T.~V. Khoa, D.~H. Son, D.~T. Hoang, N.~L. Trung, T.~T.~T. Quynh, D.~N. Nguyen, N.~V. Ha, and E.~Dutkiewicz, ``Collaborative learning for cyberattack detection in blockchain networks,'' \emph{IEEE Transactions on Systems, Man, and Cybernetics: Systems}, vol.~54, no.~7, pp. 3920--3933, Apr. 2024.

\bibitem{Schmidhuber2015}
J.~Schmidhuber, ``Deep learning in neural networks: An overview,'' \emph{Neural Networks}, vol.~61, pp. 85--117, Jan. 2015.

\bibitem{Hyndman1996}
R.~J. Hyndman and Y.~Fan, ``Sample quantiles in statistical packages,'' \emph{The American Statistician}, vol.~50, no.~4, pp. 361--365, Mar. 1996.

\bibitem{scikit}
\BIBentryALTinterwordspacing
Scikit-learn, ``Metrics and scoring: quantifying the quality of predictions,'' {Accessed:~Apr.~14,~2024}. [Online]. Available: \url{https://scikit-learn.org/stable/modules/model_evaluation.html}
\BIBentrySTDinterwordspacing

\bibitem{IERACITANO202051}
C.~Ieracitano, A.~Adeel, F.~C. Morabito, and A.~Hussain, ``A novel statistical analysis and autoencoder driven intelligent intrusion detection approach,'' \emph{Neurocomputing}, vol. 387, pp. 51--62, Mar. 2020.

\bibitem{Najari2021}
N.~Najari, S.~Berlemont, G.~Lefebvre, S.~Duffner, and C.~Garcia, ``Radon: Robust autoencoder for unsupervised anomaly detection,'' in \emph{14th International Conference on Security of Information and Networks (SIN)}, Dec. 2021.

\bibitem{Homayouni2020}
H.~Homayouni, S.~Ghosh, I.~Ray, S.~Gondalia, J.~Duggan, and M.~G. Kahn, ``An autocorrelation-based lstm-autoencoder for anomaly detection on time-series data,'' in \emph{IEEE International Conference on Big Data (Big Data)}, Dec. 2020, pp. 5068--5077.

\end{thebibliography}
	
\end{document}